\documentclass[aps,prl,twocolumn,groupedaddress]{revtex4}
\usepackage{graphicx}

\begin{document}

% Use the \preprint command to place your local institutional report
% number in the upper righthand corner of the title page in preprint mode.
% Multiple \preprint commands are allowed.
% Use the 'preprintnumbers' class option to override journal defaults
% to display numbers if necessary
%\preprint{}

\title{Observation of Anomalous Spin Segregation in a Trapped Fermi Gas}

% repeat the \author .. \affiliation  etc. as needed
% \email, \thanks, \homepage, \altaffiliation all apply to the current
% author. Explanatory text should go in the []'s, actual e-mail
% address or url should go in the {}'s for \email and \homepage.
% Please use the appropriate macro foreach each type of information

% \affiliation command applies to all authors since the last
% \affiliation command. The \affiliation command should follow the
% other information
% \affiliation can be followed by \email, \homepage, \thanks as well.
\author{X.~Du, L.~Luo, B.~Clancy, and J.~E.~Thomas}
\email[jet@phy.duke.edu]{}
%\homepage[]{Your web page}
%\thanks{}
%\altaffiliation{}
\affiliation{Duke University, Department of Physics, Durham, North
Carolina, 27708, USA}

%Collaboration name if desired (requires use of superscriptaddress
%option in \documentclass). \noaffiliation is required (may also be
%used with the \author command).
%\collaboration can be followed by \email, \homepage, \thanks as well.
%\collaboration{}
%\noaffiliation

% insert suggested keywords - APS authors don't need to do this
%\keywords{}

\date{\today}

\begin{abstract}
We report the observation of spin segregation, i.e., separation of
spin density profiles, in a trapped ultracold Fermi gas of $^6$Li
with a magnetically tunable scattering length close to zero. The
roles of the spin components are inverted when the sign of the
scattering length is reversed.  The observed density profiles are
in qualitative agreement with the spin-wave theory applied
previously to explain spin segregation in a Bose gas, but disagree
in amplitude by two orders of magnitude. The observed atomic
density profiles are far from equilibrium, yet they persist for
$\simeq$ 5 seconds in a trap with an axial frequency of $\simeq$
150 Hz. We attribute this long lifetime to Fermi statistics: The
scattering amplitude is nonzero only for atoms in opposite states,
and vanishes for atoms in the same state. By measuring the
magnetic field at which spin segregation ceases, we precisely
determine the zero crossing in the scattering length of $^6$Li as
$527.5\pm0.2$ G.
\end{abstract}


\maketitle

Ultracold Fermi gases near a  Feshbach
resonance~\cite{Feshbachpeople} exhibit strong
interactions~\cite{O'Hara2002} and offer unprecedented opportunities
to test nonperturbative quantum many-body theory in systems from
high temperature superconductors to nuclear matter. Near the
resonance, a bias magnetic field tunes the interaction between two
spin states from strongly repulsive to strongly attractive. These
strongly interacting Fermi gases have been extensively studied over
the last few years~\cite{Stringari2007}.

In contrast, the weakly interacting regime of a Fermi gas near a
Feshbach resonance has received relatively little attention. For a
bias magnetic field near the zero crossing,  the s-wave scattering
length for atoms in opposite spin states can be tuned smoothly
from  small and positive to small and negative as the bias field
is varied.

\begin{figure}[tb]
\includegraphics[width=3.5in]{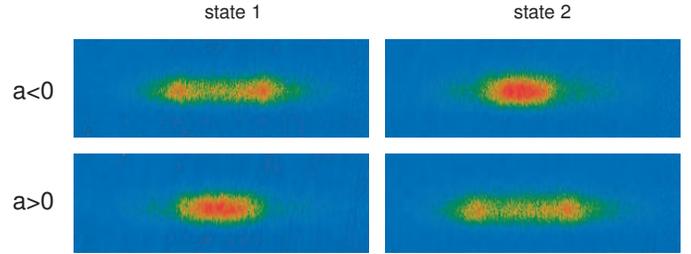}
\caption{Absorption images (state 1 and state 2) taken at 200 ms
after the RF pulse for 526.2 G (scattering length $a<0$) and for
528.8 G ($a>0$). Each image is 1.2 mm in the horizontal
direction.}
 \label{fig:ssopposite}
 \end{figure}

We report the observation of spin segregation in an
optically-trapped ultracold Fermi gas of $^6$Li with a bias magnetic
field close to the zero crossing. A sample of ultracold $^6$Li gas
is created in a coherent superposition of the two lowest spin states
$|1\rangle$ and $|2\rangle$, with equal population in each state.
Initially, the two spin states have the same atomic density profile.
After several hundred ms, the spatial densities segregate, with one
spin state moving outward and one moving inward, as shown in
Fig.~\ref{fig:ssopposite}. The roles of the spin states are
interchanged when the sign of the scattering length is reversed. The
differential trapping potential experienced by atoms in the two
states cannot account for the spin segregation, since it is four
orders of magnitude smaller than the thermal energy $k_BT$. A
longitudinal spin wave theory~\cite{Oktel2002, Williams2002,
Fuchs2002} has been used to explain spin segregation that was
observed in a trapped Bose gas~\cite{Lewandowski2002}. In the Bose
gas, numerical simulations are in good quantitative agreement with
the experimental results. In contrast, our observations of spin
segregation in a Fermi gas show great discrepancy with the
theoretical calculations based on a spin wave assumption.

In the experiments, a sample of $^6$Li atoms in a 50/50 mixture of
the two lowest hyperfine states  is loaded into a CO$_2$ laser
trap with a bias magnetic field of 840 G, where the two states are
strongly interacting. Evaporative cooling is performed to lower
the temperature of the sample~\cite{O'Hara2002}. The magnetic
field is then increased in 0.8 seconds to  a weakly interacting
regime at 1200 G where an on-resonance optical pulse of 15 $\mu$s
is applied to remove atoms of one state while leaving atoms in the
other state intact. With the single state present, the magnetic
field is lowered in 0.8 seconds to 528 G, near the zero crossing.
Then an RF pulse is applied on the $|1\rangle-|2\rangle$
transition to create spin coherence. Instead of applying a $\pi/2$
pulse,  the RF frequency is swept across the resonance of the
hyperfine transition $\sim$75.60 MHz to achieve the coherent spin
transfer. We control the amplitude of the spin transfer by varying
the sweep rate. Typically, we sweep the RF frequency by 35 kHz
(centered at hyperfine transition) in 40 ms in order to transfer
about $50\%$ of the atoms. Finally, we take absorption images of
atoms in both states (in separate experimental cycles) at various
times after the RF pulse. From those images, we can determine the
parameters of the cloud sample such as temperature and density. At
the final optical trap depth, the measured trap oscillation
frequency in the transverse directions is $\omega_\perp=2\pi\times
4360$ Hz, while the axial frequency is $\omega_z=2\pi\times 145$
Hz at 528 G. Atoms in each of the two spin states experience
slightly different magnetic trap potentials in the bias magnetic
field. Due to the bias magnetic field curvature, the differential
potential $U_1-U_2$ varies across the sample and
${d^2(U_1-U_2)/h}/{dz^2}=4.4\times10^{-4}$ Hz/$\mu$m$^2$. The
total number of atoms is $N\simeq 2.0\times 10^5$. The
corresponding Fermi temperature is $T_F\simeq 7$ $\mu$K.

We find that the spin segregation is not sensitive to the
temperature of the cloud. We observed spin segregation over the
range from a degenerate Fermi gas ($T\simeq 0.15 T_F$) to a
non-degenerate one ($T\simeq 4 T_F$). All the data shown in this
Letter were taken in the non-degenerate regime.

\begin{figure*}
\includegraphics[width=4.5in]{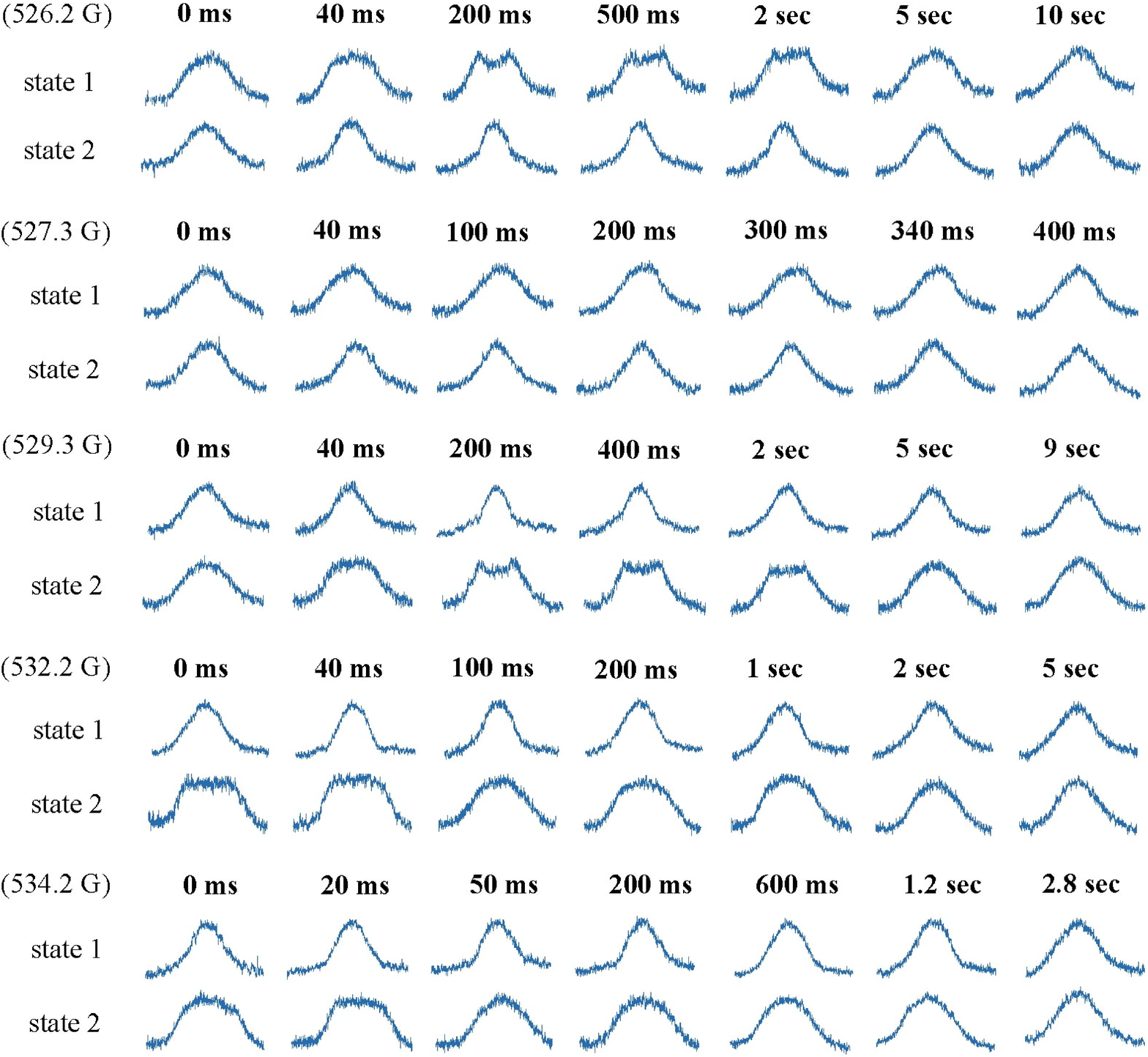}
\caption{Axial atomic density profiles (integrated over transverse
directions) at various times after the RF pulse for different
magnetic fields. The scattering length is estimated with
$a(B)=\acute{a}(B-B_0)$, where $\acute{a}=3.5$
$a_0$/G~\cite{Grimm2005, Julienne2008} and $B_0=527.5$ G
(described below) Note that different graphs have different time
scales.}
 \label{fig:ssDiff}
 \end{figure*}

The experimental results for spin segregation as a function of
bias magnetic field are summarized in Fig.~\ref{fig:ssDiff}. Here,
the sample temperature is $T\simeq 27$ $\mu$K and the peak atomic
density is $1.2\times10^{12}/cm^3$. The axial 1/e width for a fit
of a gaussian distribution to the initial density profile of the
sample is $\simeq 300 \,\mu$m.

We use the data at 529.3 G, Fig.~\ref{fig:ssDiff}, as an example
to describe the observed spin segregation in our experiments. Note
that magnetic fields are calibrated with an RF spectroscopic
technique, as described below. Spin segregation occurs only after
a coherent superposition of the two states is created. Spin
segregation starts to build up during the RF pulse duration of 40
ms, as we sweep the RF frequency across the hyperfine transition.
Starting about 40 ms after the RF pulse, we observe a significant
change in the spatial profiles of the initially overlapping clouds
for states $|1\rangle$ and $|2\rangle$. The cloud size of state
$|1\rangle$ decreases while the  density profile of state
$|2\rangle$ evolves into a double-peak structure. Maximum
segregation occurs at about 200 ms and is maintained for several
seconds. Then the two segregated clouds relax back to their
original overlapping profiles in $\sim$9 seconds.

Spin segregation observed in an ultracold Bose
gas~\cite{Lewandowski2002} has been described previously as
overdamped spin wave~\cite{Oktel2002, Williams2002, Fuchs2002}. The
basic idea can be described in terms of a Bloch vector. We take the
axis \emph{w} as the ``longitudinal'' population inversion, and the
axes \emph{u} and \emph{v} as the ``transverse'' coherence. An RF
pulse prepares a coherent superposition of spin-up and spin-down
with polarization in the \emph{u-v} plane. Due to the magnetic field
inhomogeneity, the hyperfine transition frequency varies along the
long axial direction of the sample $\sim$10 Hz. After the RF pulse,
the inhomogeneous precession of the Bloch vectors about the
\emph{w}-axis starts to build up a spin orientation gradient in
\emph{u-v} plane along the axial direction. Meanwhile, atoms with
different spins move around and collide with each other. The
interaction due to the binary collision leads to precession of each
atom's spin about the total spin vector of both atoms. Since both
atoms have spins in \emph{u-v} plane, the precession of each spin
about the total spin in \emph{u-v} plane produces spin components
\textit{out} of the \emph{u-v} plane. The subsequent formation of a
\emph{w} component leads to a spatially varying population inversion
and accounts for the spin segregation.

We tried to use a spin-wave theory to explain the spin segregation
observed in our experiment. We derived equations for the spin
density and spin current using a Wigner function operator
approach~\cite{Thomas1995}. The Wigner phase space operators are
written in terms of position space creation and annihilation
operators
\begin{equation}
\hspace*{-0.06in}\hat{W}_{\alpha\alpha'}(\mathbf{r},\mathbf{p})\equiv\int\hspace*{-0.06in}
\frac{d^3\mathbf{\epsilon}}{(2\pi\hbar)^3}
e^{\frac{i\mathbf{p}\cdot\mathbf{\epsilon}}{\hbar}}\hat{\psi}_{\alpha'}^{+}
\left(\mathbf{r}+\frac{\mathbf{\epsilon}}{2}\right)
\hat{\psi}_{\alpha}\left(\mathbf{r}-\frac{\mathbf{\epsilon}}{2}\right).
\label{wigner_operator}
\end{equation}
The Heisenberg equations are
$\dot{\hat{W}}_{\alpha\alpha'}(\mathbf{r},\mathbf{p})=
\frac{i}{\hbar}[\hat{H},\hat{W}_{\alpha\alpha'}(\mathbf{r},\mathbf{p})]$,
where $\hat{H}$ is the Hamiltonian operator
\begin{eqnarray}\label{hamiltonian}
\nonumber \hat{H} &=& \sum_{j}\int\hspace*{-0.06in}
d^3\mathbf{r}\,\hat{\psi}_j^+(\mathbf{r})
\left[-\frac{\hbar^2}{2m}\mathbf{\nabla}^2+U_j+\hbar\omega_j\right]\hat{\psi}_j(\mathbf{r})+ \\
& &\hspace*{-0.30in}\nonumber \sum_{j,k}\int\hspace*{-0.06in}
d^3\mathbf{r}_1\hspace*{-0.06in} \int\hspace*{-0.06in}
 d^3\mathbf{r}_2\,
 \hat{\psi}_j^+(\mathbf{r}_1)\hat{\psi}_k^+(\mathbf{r}_2)u_{jk}(\mathbf{r}_1-\mathbf{r}_2)\hat{\psi}_k(\mathbf{r}_2)\hat{\psi}_j(\mathbf{r}_1)
\end{eqnarray}
with
$U_j=\frac{1}{2}m\omega_\bot^2(x^2+y^2)+\frac{1}{2}m\omega_{zj}^2z^2$
and $u_{jk}(\mathbf{r}_1-\mathbf{r}_2)=\frac{1}{2}\hbar
g_{jk}\delta(\mathbf{r}_1-\mathbf{r}_2)$. Here, $j,k$ refer to
states $|1\rangle$ or $|2\rangle$; $U_j$ is the total external
potential (combined optical and magnetic) for atoms in state $j$;
$\hbar\omega_j$ is hyperfine energy of state $j$;
$g_{jk}=4\pi\hbar a_{jk}/{m}$ with $m$ thes atomic mass. $a_{jk}$
is the scattering length for atoms between state $j$ and state
$k$.

Spin density operators, $\hat{s}_x=[\hat{\psi}_{1}^{+}(\mathbf{r}
)\hat{\psi}_{2}(\mathbf{r})+H.c.]/2$, etc.,  can be written in
terms of the momentum integral of the Wigner phase space
operators,
\begin{equation}\label{spin_density_operator}
\int \hat{W}_{\alpha\alpha'}(\mathbf{r},\mathbf{p})
d^{3}\mathbf{p}=\hat{\psi}_{\alpha'}^{+}(\mathbf{r}
)\hat{\psi}_{\alpha}(\mathbf{r}).
\end{equation}
 The corresponding
spin current operators are obtained using
\begin{equation}\label{spin_current_operator}
\int
\frac{\mathbf{p}}{m}\hat{W}_{\alpha\alpha'}(\mathbf{r},\mathbf{p})
d^{3}\mathbf{p}=
-\frac{i\hbar}{2m}\hat{\psi}_{\alpha'}^{+}(\mathbf{r})
\mathbf{\nabla}\hat{\psi}_{\alpha}(\mathbf{r})+ H.c.
\end{equation}

We obtain the spin-wave equations for both a Bose gas and a Fermi
gas,
\begin{equation}\label{spin_density}
\frac{\partial S^{(i)}}{\partial
t}+\mathbf{\nabla}\cdot\mathbf{J}^{(i)}=(\mathbf{\Omega}\times\mathbf{S})^{(i)},
\end{equation}
\begin{eqnarray}\label{spin_current}
\nonumber \frac{\partial\mathbf{J}^{(i)}}{\partial
t}+\alpha\mathbf{\nabla} S^{(i)} &=&
(\mathbf{\Omega}'\times\tensor{J})^{(i)}-\gamma\mathbf{J}^{(i)} \\&
&
-\frac{1}{2m}S^{(i)}\mathbf{\nabla}U-\frac{1}{4m}n\mathbf{\nabla}U'^{(i)},
\end{eqnarray}
where
    \begin{eqnarray} \label{parameters}
    % \nonumber to remove numbering (before each equation)
     \nonumber &\mathbf{\Omega} =(\omega_0+\delta\omega_1+\delta\omega_2)\hat{z}, \\
     \nonumber &\delta\omega_1 = \frac{1}{\hbar}(U_1-U_2),\\
     \nonumber &\delta\omega_2= 2g_{11}n_1-2g_{22}n_2-2(\epsilon+1)g_{12}S^{(z)},\\
     \nonumber &\mathbf{\Omega}'=\mathbf{\Omega}+2\epsilon g_{12}\mathbf{S}, \\
     \nonumber &U=U_1+U_2+2\hbar g_{11}n_1+2\hbar g_{22}n_2+\hbar g_{12}n, \\
     \nonumber &\mathbf{U}'=\hbar\mathbf{\Omega}'.
    \end{eqnarray}
Here $i=x,y,z$ is the  spin space index: $S^{(i)}$ is a spin
density. For example, $S^{(z)}=\frac{1}{2}(n_1-n_2)$ with $n_1$,
$n_2$ the atom densities in states $|1\rangle$ and $|2\rangle$.
$\mathbf{J}^{(i)}$ is the spin current corresponding to spin
component $i$~\cite{spin_current}. $\omega_0/2\pi$ is the
hyperfine transition frequency. $\alpha=k_B T/m$. $n$ is the total
atom number. $\gamma$ is the relaxation rate, which we take to be
0 in modelling the data. Finally, $\epsilon=+1(-1)$ for Bosons
(Fermions) and $a_{11} = a_{22} = 0$ for Fermions.

%MIGHT be better to use \mathbf{v} instead of \vec{v}. This reads better and takes less vertical space in lines.

Numerical simulations based on these equations quantitatively
reproduce prior calculations of spin segregation in Bose
gases~\cite{Oktel2002, Williams2002, Fuchs2002} that agree with the
 experiment. In contrast, these equations only qualitatively describe
the spin segregation observed in our experiments with a Fermi gas.
The equations correctly predict the inversion of the roles of the
spins when the scattering length is reversed in sign. However, the
predicted amplitude of the spin segregation derived from the
calculations is two orders of magnitude smaller than what we
observe; the predicted segregation time is one order of magnitude
smaller than what we observe. This suggests that additional physics
is important for explaining spin segregation in a Fermi gas.

Fig.~\ref{fig:ssDiff} also shows the damping time (defined as the
time for the cloud to relax back to its original density profile)
increases as the scattering length is reduced in magnitude. It is
interesting to note that for certain magnetic fields (for example,
526.2 G, where the scattering length is about $5a_0$, $a_0$ is Bohr
radius), spin segregation can be maintained for up to 5 s. This
result is strikingly longer than that observed in the previous
experiment with a Bose gas, where the samples relaxed from maximum
spin segregation back to their original density profiles in about
200 ms. We attribute this result to the facts that  ultracold
fermionic $^6$Li atoms interact only via s-wave scattering between
{\it opposite} spin states and that the $^6$Li scattering length
near the zero crossing is much smaller than for the Bose gas
experiments, $\sim5a_0$ compared to $\sim100a_0$ for $^{87}$Rb.

We also  note that the decoherence time in our experiment is
measured to be about 70 ms, which is comparable to the longest
spin segregation time that we observe. The decoherence time is the
measured time scale for the phase of the coherent superposition to
decay, producing a simple binary mixture.  Decoherence is caused
by the spatial variation of the magnetic field across the sample.
The short decoherence time may explain why we can not clearly
observe spin segregation when the magnetic field is too close to
the zero crossing. When the scattering length is very small, the
time required for spin segregation to build up is longer than the
decoherence time.

An additional experiment was performed to further study the spin
segregation. After we create a sample that reaches the maximum
spin segregation in 200 ms at 529 G, we immediately blow out atoms
of one state using a resonance optical pulse. Then we take images
of the remaining state at various times after the optical pulse.
We observe that spin segregation maintains its atomic density
profile for $\sim2$ seconds. This shows that slow relaxation of
one state does not depend on presence of the other state. As we
know, a single-peak gaussian distribution is a stationary
equilibrium state for a thermal cloud in a harmonic trap. In
contrast, the double-peak and one-peak non-gaussian atomic density
profiles observed in our experiments are far from equilibrium. It
is remarkable that we can create a non-equilibrium stationary
system that has such a long lifetime, $\sim2$ s, compared to the
axial trap period  6.9 ms and segregation time $\sim200$ ms.
\begin{figure}[tb]
\includegraphics[width=3.5in]{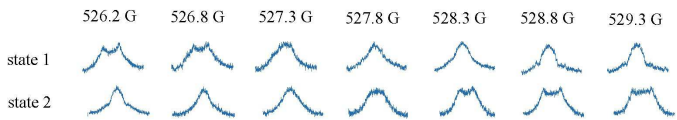}
\caption{Axial atomic density profiles at 200 ms after coherence
is created by an RF pulse, for different magnetic fields. The zero
crossing point in the scattering length is determined by studying
the [transition--NOT a transition] in the spin segregation
behavior as the magnetic field is varied.}
 \label{fig:zero_crossing}
 \end{figure}

Previous experiments have shown that a zero crossing occurs in the
$|1\rangle-|2\rangle$ scattering length of $^6$Li near 528
G~\cite{O'Hara2002_2,Grimm2002}. We now describe a new
determination of the zero crossing based on the behavior of the
spin segregation as the magnetic field is varied. As described
above, the roles of the spin states are interchanged as the
scattering length $a$ changes sign from negative to positive.
Fig.~\ref{fig:ssDiff} shows data for $a<0$ at 526.2 G and for
$a>0$ at 529.3 G. The spin segregation time (defined as the time
for the cloud to reach maximum spin segregation) decreases
(increases) as the scattering length increases (decreases). Spin
segregation becomes difficult to observe when the magnetic field
is too close to the zero crossing, for example, 527.3 G in the
figure, as the segregation time then far exceeds the decoherence
time, as noted above. Similarly, very far from the zero crossing,
$\sim$15 G above or below collisional relaxation dominates and
spin segregation is not observable (not shown in the figure).

Taking advantage of the fact that spin segregation slows for
magnetic fields very close to the zero crossing, we determine the
zero crossing precisely. We take absorption images of both states
at 200 ms after the RF pulse for different magnetic fields.
Fig.~\ref{fig:zero_crossing} shows a clear change in the spin
segregation behavior from 526.2 G to 529.3 G. The magnetic fields
are calibrated by RF spectroscopy using the Hamiltonian for the
$^6$Li ground state hyperfine interactions in a magnetic field,
\begin{equation}
H_{int}=\frac{a_{hf}}{\hbar^2}\mathbf{S}\cdot\mathbf{I}-\frac{\mu_B}{\hbar}(g_J\mathbf{S}+g_I\mathbf{I})\cdot\mathbf{B},
\label{eq:hyperfine}
\end{equation}
where $\mu_B/h=1.399624604$ MHz is the Bohr magneton,
$a_{hf}/h=152.1368407$ MHz~\cite{hyperfine_constant} is the ground
state magnetic hyperfine constant for $^6$Li,
$g_J=-2.0023010$~\cite{hyperfine_constant} is the total electronic
g-factor for the $^6$Li ground state,
$g_I=0.0004476540$~\cite{hyperfine_constant} is the nuclear
g-factor, $h$ is Planck's constant, and $\mathbf{B}$ is the bias
magnetic field.

The transition in the  behavior of the segregation occurs at 527.5
G, which is determined as follows. We fit the axial atomic density
profiles of both states with a gaussian distribution for the data
taken at 527.3 G and 527.8 Gs. For each data set, we derive
gaussian widths of both clouds and their ratio. The ratio of the
width of state $|1\rangle$ to that of state $|2\rangle$ is greater
than 1 for 527.3 G; less than 1 for 527.8 G. By interpolation, we
find the point where the ratio of widths is equal to 1, which
determines the value of magnetic field
 $527.5\pm0.2$ G (the error bar is the quadratic combination of
 the shot-to-shot fluctuation of the magnetic field setting and the
  magnetic field calibration error arising from the RF
spectrum fit). Since no spin segregation occurs with the
scattering length of zero, we interpret  the value of magnetic
field where the ratio of widths is 1 to be the zero crossing.

In conclusion, we have observed spin segregation in a trapped
ultracold atomic Fermi gas of $^6$Li with a scattering length
close to zero. The observed behavior of the spin segregation
appears to require a modification of the current theory based on
spin waves, or possibly a new mechanism. In addition, using spin
segregation, we have precisely determined the zero crossing in the
scattering length of $^6$Li.

%Can we relate our results and equations to Zwierlein's no spectral shift paper?

This research is supported by ARO, NSF, and DOE.


\begin{references}
\bibitem{spin_current}$\tensor{J}$ is a tensor and $\mathbf{J^{{i}}}$
is a vector in coordinate space.
\end{references}
 \end{document}